\documentclass[prb,twocolumn,showpacs]{revtex4}
\usepackage{graphicx}% Include figure files
\usepackage{dcolumn}% Align table columns on decimal point
\usepackage{bm}% bold math
\begin{document}

\preprint{APS/123-QED}

\title{Finite-temperature Mott transitions in multi-orbital Hubbard model}
% Force line breaks with \\

\author{Kensuke Inaba}
% \altaffiliation[Also at ]
%{Department of Applied Physics, Osaka University, Suita, Osaka 565-0871, Japan.}
%Lines break automatically or can be forced with \\
\author{Akihisa Koga}%
\author{Sei-ichiro Suga}
\author{Norio Kawakami}
% \email{Second.Author@institution.edu}
\affiliation{%
Department of Applied Physics, Osaka University, Suita, Osaka 565-0871, Japan
}%

% \homepage{}
%\affiliation{
%Second institution and/or address\\
%This line break forced% with \\
%}%

\date{\today}% It is always \today, today,
             %  but any date may be explicitly specified

\begin{abstract}
We investigate the Mott transitions in the multi-orbital Hubbard model 
at half-filling by means of the self-energy functional approach.
The phase diagrams are obtained at finite temperatures
for the Hubbard model with up to four-fold degenerate bands. 
We discuss how the first-order Mott  transition 
 points $U_{c1}$ and $U_{c2}$ as well as
 the critical temperature $T_c$ depend on the orbital degeneracy.
It is elucidated that enhanced orbital fluctuations play a key 
role to control the  Mott transitions in  the multi-orbital
Hubbard model.
\end{abstract}

\pacs{71.30.+h, 71.10.Fd}% PACS, the Physics and Astronomy
                             % Classification Scheme.
%\keywords{Suggested keywords}%Use showkeys class option if keyword
                              %display desired
\maketitle
%%%%%%%%%%%%%%%%%%%%%%%%%%%%%%%%%%%%%%%%%%%%%%%%%%%%%%%%%%%%%%%%%%%%
%%%                          introduction                         %%
%%%%%%%%%%%%%%%%%%%%%%%%%%%%%%%%%%%%%%%%%%%%%%%%%%%%%%%%%%%%%%%%%%%%
%----------------------------   intro 1   ---------------------------

%%%%%%%%%%%%%%%%%%%%%%
\section{Introduction}
%%%%%%%%%%%%%%%%%%%%%%

The Mott transition has been one of the most attractive topics 
in strongly correlated electron systems.\cite{Mott,Gebhard} 
There are a number of prototype materials in transition 
metal oxides. A well-known example is 
$\rm V_2O_3$,\cite{V2O3,ImadaRev} which
exhibits the first-order metal-insulator transition 
at a certain critical temperature. 
The phase diagram obtained by systematic experiments is consistent with 
theoretical studies of the single orbital Hubbard model 
by means of dynamical mean field theory
(DMFT).\cite{GeorgesRev,KotliarPT}  This implies that the DMFT treatment 
of the Hubbard model, which properly takes 
into account local electron correlations, 
captures the essence of the Mott transition.
However, in order to give more quantitative discussions 
about the Mott transitions, it is
indispensable to incorporate the effect of the 
 orbital degeneracy, which 
often gives rise to rich phase diagrams, as observed for
$\rm La_{1-x}Sr_xMnO_3$,\cite{TokuraScience} 
$R \rm TiO_3$\cite{Katsufuji97,Mochizuki03}, etc.
More recently,  the orbital-selective Mott transition 
has attracted considerable attention,
\cite{Anisimov02,Sigrist04,Liebsch,KogaOS,Ferrero05,Medici05,Arita05} 
in connection with 
the materials such as
$\rm Ca_{2-x}Sr_xRuO_4$\cite{Nakatsuji} and 
$\rm La_{n+1}Ni_nO_{3n+1}$\cite{LaNiO,Kobayashi96}.

These interesting experimental findings have stimulated 
a number of theoretical works on 
the Mott transitions in the two-orbital Hubbard model
by means of DMFT.\cite{Kotliar96,Rozenberg97,Bunemann:Gutzwiller,Hasegawa98,Held98,Han98,Momoi98,Klejnberg98,Imai01,Koga:ED_DMFT,OnoDP,Ono03,Pruschke04,Liebsch,KogaOS,Ferrero05,Medici05,Arita05}
Although the two-orbital model has been studied 
in detail, a systematic study on the systems
having  more orbitals is still lacking at finite temperatures.  
For example, the finite-temperature Mott transitions 
for the multi-orbital Hubbard models
has not been investigated quantitatively  by DMFT. One of the 
difficulties lies in the practical calculation within the DMFT framework.
Quantum Monte Carlo simulations, which can be 
a powerful numerical method to treat local correlations in DMFT,
encounter sign problems. Also, Wilson's renormalization group method
gets more difficult to apply as the number of orbitals increases.
A much more simplified approach, the two-site DMFT,\cite{Potthoff01}  
is not efficient enough to treat finite-temperature properties.
It is thus desirable to systematically investigate electron correlations
in the multi-orbital Hubbard model at finite temperatures.

Motivated by the above hot topics, we consider the Mott transitions
in the multi-orbital Hubbard model at zero as well as
finite temperatures.  For this purpose, 
we make use of a self-energy functional approach (SFA)
proposed by Potthoff recently.\cite{SFA,Potthoff04}
  This method, which 
is based on the Luttinger-Ward functional approach,\cite{Luttinger}
gives a powerful tool to discuss electron correlations.
A remarkable point is that this approach provides an
 efficient way to deal with finite-temperature properties 
of the multi-orbital
Hubbard model.  The main purpose of the present paper 
is to determine the finite-temperature
phase diagram of the Hubbard model with up to four-fold bands, and 
 quantitatively discuss how the orbital degeneracy affects 
 the Mott transitions.
 
The paper is organized as follows. In the next section,
we introduce the model Hamiltonian and
briefly summarize the formulation of SFA.
In Sec. \ref{sec:two_band} we 
present the detailed analysis based on SFA by exploiting the two-orbital 
model as a prototype system. Then in Sec.  \ref{sec:multi_band}
we determine the phase diagram, and 
discuss the nature of the Mott transitions in multi-orbital systems in detail.
A brief summary is given in Sec. \ref{sec:summary}.

%%%%%%%%%%%%%%%%%%%%%%%%%%%%%%%%%%%%%%%%%%%%%%%%%%%%%%%%%%%%%%%%%%%%
%%%                             model                             %%
%%%%%%%%%%%%%%%%%%%%%%%%%%%%%%%%%%%%%%%%%%%%%%%%%%%%%%%%%%%%%%%%%%%%

\section{Model and Method}\label{sec:model_method}

We consider a correlated electron 
system having $M$ degenerate orbitals, which is described by
the following multi-orbital Hubbard Hamiltonian,
%%%%%%%%%%%%%%%%%%%%%%%%%%%%%%%
\begin{eqnarray}
{\cal H}&=&{\cal H}_0+{\cal H}^\prime,\\
{\cal H}_0&=&\sum_{<i,j>}\sum_{\alpha=1}^M \sum_{\sigma}t_\alpha
          c^\dag_{i\alpha\sigma} c_{j\alpha\sigma},\\
{\cal H}^\prime&=&
            U \sum_i\sum_{\alpha=1}^M
                n_{i \alpha \uparrow} n_{i \alpha \downarrow}
                \nonumber\\
          &+&   U^\prime \sum_i\sum_{\alpha<\alpha^\prime=1}^M
                \sum_{\sigma \sigma^\prime}
                n_{i \alpha \sigma} n_{i \alpha^\prime \sigma^\prime}
%                \nonumber\\
%          &-&   J \sum_{\alpha<\alpha^\prime=1}^M
%                \sum_{\sigma }
%                n_{i \alpha \sigma} n_{i \alpha^\prime \sigma}
%                \nonumber\\
%          &-&   J \sum_{\alpha<\alpha^\prime=1}^M
%                (c^\dag_{i \alpha \uparrow}c_{i \alpha \downarrow}
%                c^\dag_{i \alpha \downarrow}c_{i \alpha \uparrow}+H.c.)
                \label{eq:original_model},
\end{eqnarray}
%%%%%%%%%%%%%%%%%%%%%%%%%%%%%%%%%%%%%%%%%%%%%
where $c^\dag_{i\alpha\sigma}(c_{i\alpha\sigma})$ creates (annihilates) 
an electron with spin $\sigma (=\uparrow, \downarrow)$ and orbital
$\alpha (=1,2,3\cdots ,M)$ at the $i$ th lattice site, 
%We set in the paper $J=0$ and $U=U^\prime$ for simplicity.
%We imposed a condition of $U=U^\prime+2J$ for the symmetry of 
%degenerate orbital. 
$t_\alpha$ denotes the hopping integral for orbital $\alpha$, $U (U')$ 
is the intra-orbital (inter-orbital) Coulomb interaction.
%, and $J$ the Hund coupling.
% (DOS) 
%$\rho(\omega)=(\pi/2D)\sqrt{1-(\omega/D)^2}$, 
%where $D$ is half of the band width, which is realized on the Bethe lattice. 
In the following, we mainly consider the case of $U=U'$, and
give brief discussions for more general cases including the Hund coupling
in the end of the paper.

%%%%%%%%%%%%%%%%%%%%%%%%%%%%%%%%%%%%%%%%%%%%%%%%%%%%%%%%%%%%%%%%%%%%
%%%                             method                            %%
%%%%%%%%%%%%%%%%%%%%%%%%%%%%%%%%%%%%%%%%%%%%%%%%%%%%%%%%%%%%%%%%%%%%
%===================================================================
%==                         outline of SFA                        ==
%===================================================================

In order to address the  Mott transitions
in the multi-orbital Hubbard model, we use SFA.\cite{SFA,Potthoff04}
We briefly summarize the essence of SFA.
For a correlated electron system, the grand potential is 
generally expressed as 
%%%%%%%%%%%%%%%%%%%%%%%%%%%%%%%%%%%%%%%%%
\begin{eqnarray}
\Omega [{\boldsymbol \Sigma}]&=&F[{\boldsymbol \Sigma}]+
{\rm Tr}\ln [-({\bf G}_0^{-1}-{\boldsymbol \Sigma})^{-1}]\label{eq:LW},
\end{eqnarray}
%%%%%%%%%%%%%%%%%%%%%%%%%%%%%%%%%%%%%%
where $F[{\boldsymbol \Sigma}]$ is the Legendre transformation of 
the Luttinger-Ward potential $\Phi [{\bf G}]$, and
${\bf G}$ (${\bf G}_0$) and ${\boldsymbol \Sigma}$ are the 
full  (bare) Green's function  and the self-energy, respectively.
The condition imposed on the functional (\ref{eq:LW}),
%%%%% %%%%%%%%%%%%%%%%%%%%%%%%%%%%%%%%%%%%%%
\begin{eqnarray}
\frac{ \partial \Omega[{\boldsymbol \Sigma}] }{ \partial {\boldsymbol \Sigma}}
=0,\label{variation}
\end{eqnarray}
%%%%%%%%%%%%%%%%%%%%%%%%%%%%%%%%%%%%%%%%%%%
gives the Dyson equation ${\bf G}^{-1}={\bf G}_0^{-1}-{\boldsymbol \Sigma}$.
An important point is that the functional form of $F[{\boldsymbol \Sigma}]$ 
does not depend on the detail of the Hamiltonian ${\cal H}_0$
 as far as the interaction term
${\cal H}^\prime({\bf U})$ keeps its shape unchanged.
This fact allows us to introduce a proper reference system having
 the same interaction term, which we denote as
${\cal H}_{\rm ref}={\cal H}_0({\bf t}^\prime)+{\cal H}^\prime({\bf U})$.
Then the grand potential is written  as,
%${\boldsymbol \Sigma}({\bf t}^\prime)$ 
%%%%%%%%%%%%%%%%%%%%%%%%%%%%%%%%%%%%
\begin{eqnarray}
\Omega [{\boldsymbol \Sigma}({\bf t}^\prime)]&=&\Omega({\bf t^\prime})
\nonumber\\
    &+&{\rm Tr}\ln
    \left[-({\bf G}_0({\bf t})^{-1}-{\boldsymbol \Sigma}
({\bf t^\prime}))^{-1}\right]
    \nonumber\\
    &-&{\rm Tr}\ln
    \left[-({\bf G}_0({\bf t}^\prime)^{-1}-{\boldsymbol \Sigma}
({\bf t^\prime}))^{-1}\right],\label{eq:omega_SFA}
\end{eqnarray}
%%%%%%%%%%%%%%%%%%%%%%%%%%%%%%%%%%%%%
where $\Omega({\bf t}^\prime)$ and ${\boldsymbol \Sigma}({\bf t}^\prime)$ are 
the grand potential and the self-energy for the reference system, whose
 bare Green's function is denoted as
 ${\bf G}_0({\bf t'})^{-1}=\omega+\mu-{\bf t'}$
($\mu$: chemical potential).
The variational condition (\ref{variation}) is rewritten as
%%%%%%%%%%%%%%%%%%%%%%%
\begin{equation}
\frac{\partial \Omega [{\boldsymbol \Sigma}
({\bf t}^\prime)]}{\partial {\bf t}^\prime}=0. 
\label{eq:variational_condition}
\end{equation}
%%%%%%%%%%%%%%%%%%%%%%%%%%%%%
By choosing the  parameters ${\bf t}^\prime$ for the reference 
system so as to
 satisfy the condition (\ref{eq:variational_condition}),
we can find the best system within the reference Hamiltonian,
 which can approximately describe the original correlated system.

It should be noticed here that SFA 
provides us with an efficient and tractable way to deal with 
finite-temperature properties of the multi-orbital
system, where standard DMFT 
combined with numerical techniques
faces difficulties in a practical computation when 
the number of orbitals  increases.  
Another notable point in this approach is that the critical behavior
 can be discussed more precisely
than the DMFT analysis,\cite{GeorgesRev} 
when one chooses the same type of the effective impurity model. 
In fact, by comparing the results obtained by the two-site 
DMFT,\cite{Bulla00,Potthoff01} 
the critical point for a single orbital Hubbard model 
obtained by SFA with the two-site model \cite{SFA}
is in good agreement with that obtained by 
DMFT with the aid of  numerical 
techniques.\cite{Caffarel94,Moeller95,Bulla99}

%===================================================================
%==                 detail of method in this paper                ==
%===================================================================

To discuss the multi-orbital Hubbard model (\ref{eq:original_model}),
we exploit the Anderson impurity model as a reference system
in SFA.\cite{SFA,Potthoff04}
The Hamiltonian for the reference system is explicitly given as 
%%%%%%%%%%%%%%%%%%%%%%%%%%%%%
\begin{eqnarray}
{\cal H}_{\rm ref}&=&\sum_i{\cal H}_{\rm ref}^{(i)},\\
%---------------    Hamiltonian of reference system    --------------
  {\cal H}_{\rm ref}^{(i)}&=&\sum_{\alpha \sigma } \varepsilon^{(i)}_{0\alpha}
          c^\dag_{i\alpha\sigma} c_{i\alpha\sigma}+\sum_{k=1}^{N_b}\sum_{\alpha \sigma} 
          \varepsilon^{(i)}_{k \alpha}
          a^{(i)\dag}_{k\alpha\sigma}a^{(i)}_{k\alpha\sigma}\nonumber\\
          &+&\sum_{k=1}^{N_b}\sum_{\alpha \sigma}V^{(i)}_{k \alpha}
          (c^\dag_{i\alpha\sigma}a^{(i)}_{k\alpha\sigma}+H.c.)\nonumber\\
          &+&U \sum_{\alpha}
                n_{i \alpha \uparrow} n_{i \alpha \downarrow}
                +U^\prime \sum_{\alpha<\alpha^\prime}
                \sum_{\sigma \sigma^\prime}
                n_{i \alpha \sigma} n_{i \alpha^\prime \sigma^\prime},\label{eq:ref_model}
\end{eqnarray}
%%%%%%%%%%%%%%%%%%%%%%%%%%%%%%%%%%
where $a^{(i)\dag}_{k\alpha\sigma}(a^{(i)}_{k\alpha\sigma})$ 
creates (annihilates) an electron with $\sigma$ spin and $\alpha$ orbital 
at the $k(=1,2,\cdots N_b)$ th site, 
which is connected to the $i$ th site in the original lattice. 
Since we consider the Mott transitions without symmetry breaking,
we set $\varepsilon_{k\alpha}$ and $V_{k\alpha}$ site-independent. 
Note that  in the limit of $N_b\to \infty $,
Eq. (\ref{eq:variational_condition}) is equivalent to
 the self-consistent equation in DMFT.\cite{SFA}  Since
the Green's function and the self-energy are diagonal with
respect to the  site, spin, and orbital indices, 
the grand potential per site reads
%%%%%%%%%%%%%%%%%%%%%%%%%%%%%%%
\begin{eqnarray}
\Omega /L&=&\Omega_{\rm imp}\nonumber\\
    &-&2 \sum _{\alpha} \int d \omega f(\omega) 
    R_\alpha \left[\omega + \mu - \Sigma_\alpha(\omega)\right]
    \nonumber\\
    &+&2 \sum_{\alpha} \sum_{k=0}^{N_b} \int d \omega f( \omega ) 
    \theta \left[ G^\prime_{k\alpha}(\omega) \right]\label{eq:Omega_DIA},\\
%\end{eqnarray}
%and
%\begin{eqnarray}
G^\prime_{0\alpha}(\omega)&=&[\omega+\mu-\varepsilon_{0\alpha}
           -\Delta_\alpha(\omega)-\Sigma_{\alpha}(\omega)]^{-1},\\
G^\prime_{k\alpha}(\omega)&=&(\omega+\mu-\varepsilon_{k\alpha})^{-1},\ \ 
(k=1,2,\cdots,N_b),\\
\Delta_\alpha(\omega)&=&\sum_{k=1}^{N_b} V_{k\alpha}^2G^\prime_{k\alpha}
(\omega),
\end{eqnarray}
%%%%%%%%%%%%%%%%%%%%%%%%%%%%%%%%%%%%%%%%%
where  $R_\alpha(\omega)=\int_{-\infty}^\omega \rho_{\alpha}(z) dz$,
%$f(\omega)$ is the Fermi distribution function 
%$\rho_\alpha(z)$ is a free density of state,
$f(\omega)=1/(1+e^{-\beta\omega})$
 and $\theta(\omega)$ is a step function. 
The grand potential and the self-energy for an 
impurity system ${\cal H}_{\rm ref}^{(i)}$ 
are denoted as $\Omega_{\rm imp}$ and $\Sigma_\alpha(\omega)$.

In the following, we focus on the half-filled Hubbard model
($\mu=(M-1/2)U$) to discuss how the orbital degeneracy $(M=1, 2, 3, 4)$
affects the Mott transitions. We take the simplest reference system $(N_b=1)$,
where we fix the parameters $(\varepsilon_{0\alpha}, \varepsilon_{1\alpha},
V_{1\alpha})=(0, \mu, V)$ for each orbital $\alpha$. 
%to treat the effect of orbital degeneracy systematically,
%we adopt this minimal reference system 
%which may be less accurate than 
%If we adopts the systems have large degree of freedom,
%we may do more accurate calculation.\cite{Pozgajcic04}
This simplified treatment is somewhat analogous to the two-site 
DMFT at zero temperature. However, the present 
approach enables us to treat finite-temperature
quantities systematically, which is a notable advantage 
beyond the two-site DMFT. 
The condition (\ref{eq:variational_condition}) is now reduced to  
$ \partial \Omega/ \partial V =0$.
We note that the hopping integral $V$ between a given site on the
original  lattice and a fictitious site corresponds, roughly speaking, to
the renormalized bandwidth of quasi-particles.
In fact, Fermi-liquid properties at zero temperature are determined by 
the value of $V$. For instance, 
small $V$ implies the formation of heavy quasi-particles,
and  the system enters the insulating phase at $V=0$.
By calculating the effective hybridization $V$, 
we thus discuss the stability of the metallic state in 
the multi-orbital systems.

%%%%%%%%%%%%%%%%%%%%%%%%%%%%%%%%%%%%%%%%%%%%%%%%%%%%%%%%%%%%%%%%%%%%
%%%                        two-band system                        %%
%%%%%%%%%%%%%%%%%%%%%%%%%%%%%%%%%%%%%%%%%%%%%%%%%%%%%%%%%%%%%%%%%%%%

\section{Detail of calculations: two-orbital system}\label{sec:two_band}

Let us start with the two-orbital Hubbard model, by which we
illustrate the basic procedure of SFA to determine the 
phase diagram for more general  multi-orbital cases.
We adopt here a semielliptic density of states 
$\rho (\varepsilon)=(\pi/W)\sqrt{1-(2\varepsilon /W)^2}$ 
with the bandwidth $W=4$.

%At zero temperature, it was pointed out that orbital fluctuations 
%induced by the additional orbital degrees of freedom are important
%to understand the Mott transitions in the system.
%However, the role of orbital fluctuations at finite temperatures 
%has not been clarified yet.
%In the realistic system, 
%it may be important to understand the Mott transitions% 
%at finite temperatures. Therefore, it is desirable to discuss it.
%/* introduction ?
%In contrast to the single band Hubbard model, 
%the addition of the orbital degrees of freedom may lead to orbital
%fluctuations, inducing interesting phenomena such as 
%orbital frustration and orbital selective Mott transition.
%%%%%%%%%%%%%%%%%%
We first look at the ground-state properties by
examining the stationary points in the grand potential.
%%%%%%%%%%%%%%%%%%%%%%%%%%%%%%%%%%%%%%%%%%%%%%%%%%%%%%%%%%%%%%%%%
\begin{figure}[htb]
\includegraphics[width=\linewidth]{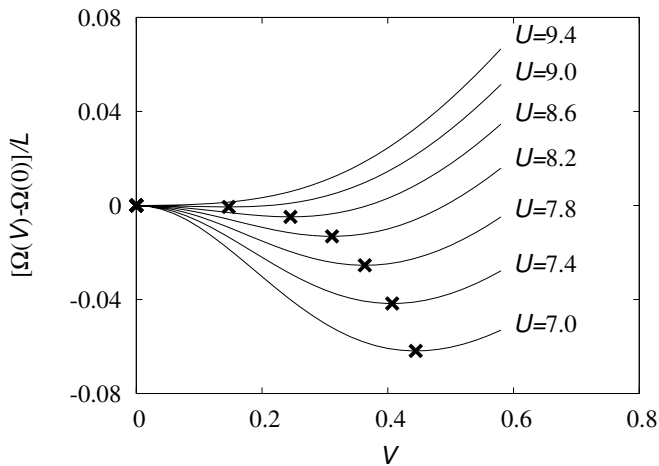}
\caption{The grand potential $\Omega(V)$ for the two-orbital 
model ($M=2$) as a function of $V$
for different $U$ at $T=0$.
 Crosses correspond to stationary points.}
\label{fig:omg_zeroT}
\end{figure}
%%%%%%%%%%%%%%%%%%%%%%%%%%%%%%%%%%%%%%%%%%%%%%%%%%%%%%%%%%%%%%%%%
In Fig. \ref{fig:omg_zeroT}, we show the grand potential at zero temperature
for several values of the Coulomb interaction $U$.
When $U$ is small, the grand potential has a minimum at finite $V$.
Thus the effective bandwidth is finite, 
stabilizing  the metallic ground state. 
 It is seen that the stationary point
moves towards the origin continuously with increasing $U$.
For large $U$, the minimum is located at the origin, 
indicating that the Mott insulating state is stabilized
by strong electron correlations.
In the vicinity of the critical point $U_c$, 
the grand potential can be expanded in powers of $V$,
%%%%%%%%%%%%%%%%%%%%%%%%%%%%%%%%%%%%%%%%
\begin{equation}
\Omega(V)=\Omega(0)+A V^2+{\cal O}(V^4)\label{eq:omg_expand}.
\end{equation}
%%where $A=\partial^2 \Omega/\partial V^2|_{V=0}$.
%%%%%%%%%%%%%%%%%%%%%%%%%%%%%%%%%%%%%
Therefore, the critical point separating the metallic and 
the insulating phases is characterized by the condition $A=0$.\cite{SFA}
By solving this analytically,
we obtain the self-consistent equation for the critical point $U_c$ as,
%-------------------------   analytic form   ------------------------
\begin{eqnarray}
U_c&=&-\frac{100}{39}\int^0_{-\infty} \left[z-\xi(z,U_c)\right] \rho (z)dz\nonumber\\
&-&\frac{U_c}{39}\int^\infty_{-\infty}\frac{317U_c+83 \xi(z,U_c)}{[17U_c+8 \xi(z,U_c)]\xi(z,U_c)}\rho(z)dz
\label{eq:SCEq_Uc}
\end{eqnarray}
%%%%%%%%%%%%%
with  $\xi(z,U)=\sqrt{U^2+z^2}$.
The detail of the derivation is shown in Appendix A. 
The critical point $U_c \simeq 9.217$ thus obtained 
 is more accurate than the results $(U_c=10)$
estimated by the two-site DMFT method.\cite{Koga02}

\begin{figure}[tHb]
\includegraphics[width=\linewidth]{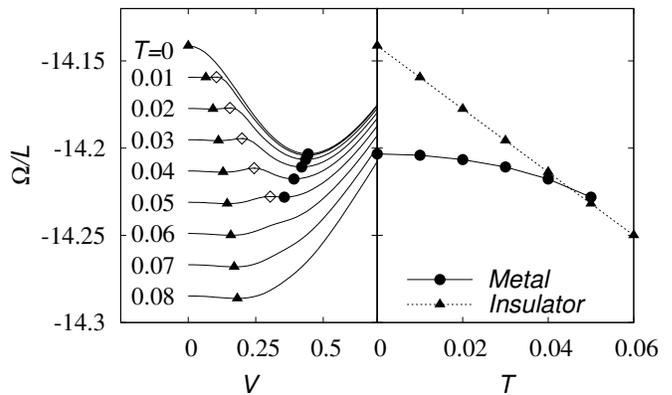}
\caption{{\it (left panel)} 
The grand potential $\Omega$ as a function of the effective 
hybridization $V$ for $U=7$. 
Closed circles and triangles represent the stationary points for 
the metallic and the Mott insulating states.
Diamonds denote unstable stationary points.
{\it (right panel)} Stationary values of
the grand potential as a function of the temperature $T$ for $U=7$. 
Symbols are the same as in the left panel. 
}\label{fig:omg_finiteT}
\end{figure}

\begin{figure}[htb]
\includegraphics[width=\linewidth]{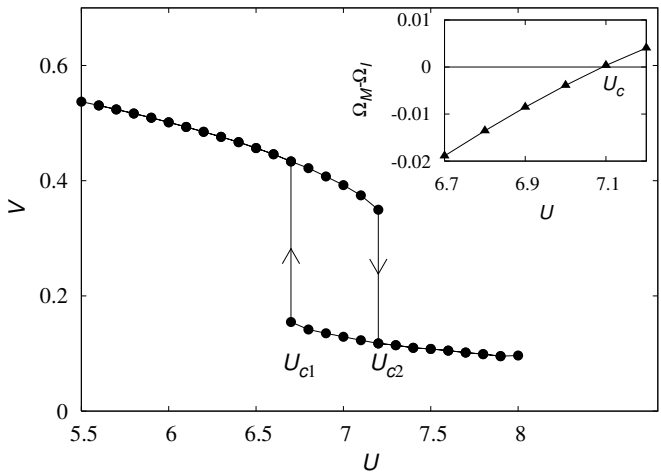}
\caption{ The effective hybridization $V$
as a function of $U$ at $T=0.04$.
Inset shows the grand potential as a 
function of $U$ at $T=0.04$, where $\Omega_M$ ($\Omega_I$) is 
the grand potential for the metallic (insulating) state. 
}
\label{fig:hys}
\end{figure}

We now consider the competition between the metallic and the Mott 
insulating phases at finite temperatures, following the way
for the single band case.\cite{SFA}
The grand potential is shown in Fig. \ref{fig:omg_finiteT} 
at finite temperatures. 
%%%%%%%%%%%%%%%%%%%%%%%%%%%%%%%
%%%%%%%%%%%%%%%%%%%%%%%%%%%%%%%%%%%%%
It is seen that two minima appear at low temperatures. 
One of the minima is located at larger $V$, which corresponds to
 the metallic solution, since it is continuously 
connected to the metallic one  at zero temperature.
The other is adiabatically connected to $V=0$ at zero temperature,
so that this solution characterizes the Mott insulating state.
Such a double-well structure at low temperatures causes
the first-order transition accompanied by hysteresis.
In fact, as increasing temperature, the stationary point for 
the metallic state 
disappears around $T_{c2}=0.05$, 
where the Mott transition occurs to the insulating phase.
On the other hand, as decreasing temperature,
the Mott insulating phase realized at high temperatures is stable
except for zero temperature, since the corresponding
local minimum always exists 
at finite temperatures.
The first-order transition temperature $T_c=0.045$ for $U=7$ is determined 
by the crossing point of the two minima in the grand potential, 
as shown in Fig. \ref{fig:omg_finiteT}.
This hysteresis induced by the first-order transition 
is also observed when the interaction $U$ is varied. 
We show the effective hybridization $V$ at $T=0.04$ in Fig. \ref{fig:hys}.
%%%%%%%%%%%%%%%%%%%%
%%%%%%%%%%%%%%%%%%%%%%%%%%%%%%%
Starting from the metallic state,
 the increase of the Coulomb interaction 
decreases the effective hybridization $V$, and
finally triggers the Mott transition to the insulating phase 
 at $U_{c2}=7.2$, where we observe the discontinuity in $V$.
On the contrary, the Mott transition occurs at $U_{c1}=6.7$ 
when the interaction decreases. 
In this parameter regime, the level crossing in the minima 
of the grand potential appears at $U_c=7.1$ (inset of Fig. \ref{fig:hys}), 
which defines the first-order transition point at $T=0.04$.

We have seen here that  the two-orbital Hubbard model exhibits
qualitatively similar properties to the single-band case \cite{GeorgesRev,SFA}
as far as the nature of  the Mott transitions is concerned.
In the following section, we give more quantitative 
discussions about the Mott transitions by using
 the phase diagrams obtained for the multi-orbital
Hubbard model with  $M=1 \sim 4$.

%%%%%%%%%%%%%%%%%%%%%%%%%%%%%%%%%%%%%%%%%%%%%%%%%%%%%%%%%%%%%%%%%%%%
%%%                      multi-orbital system                     %%
%%%%%%%%%%%%%%%%%%%%%%%%%%%%%%%%%%%%%%%%%%%%%%%%%%%%%%%%%%%%%%%%%%%%

\section{Phase diagrams of multi-orbital systems}\label{sec:multi_band}

\begin{figure}[htb]
\includegraphics[width=\linewidth]{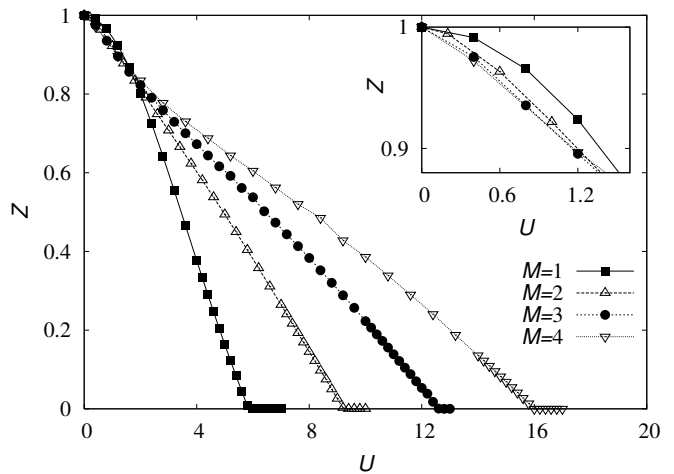}
\caption{
The quasi-particle weight $Z$  
as a function of the Coulomb interaction $U$
for the system with $M=1,2,3,4$ at $T=0$. 
The inset is the enlarged figure 
in the small $U$ region.
}\label{fig:zeroT}
\end{figure}

We  now determine the phase diagrams of
 the multi-orbital Hubbard model.
We first calculate the renormalization factor 
$Z=(1-\partial \Sigma(\omega)/\partial \omega)^{-1}$,
which is inversely proportional to the effective mass,
to characterize the metallic ground state at zero temperature.
The results are shown in Fig. \ref{fig:zeroT}. 
%%%%%%%%%%%%%%%%%%%%%%%%%%%%%%%%%%%%%%%%%%%%%%%%%%%%%%%%%%%%%%%%%
%%%%%%%%%%%%%%%%%%%%%%%%%%%%%%%%%%%%%%%%%%%%%%%%%%%%%%%%%%%%%%%%%
The introduction of the Coulomb interaction decreases 
the renormalization factor, making the mass of quasi-particles heavier. 
In the small $U$ region, there is not a big difference in the 
behavior of $Z$ among different $M$ cases.
If we look at $Z$ in more detail (inset of Fig. \ref{fig:zeroT}), 
the mass gets slightly heavier (smaller $Z$) as $M$ increases.
This comes from the fact that the electron correlations are
somewhat enhanced for multi-orbital cases by
 the inter-orbital Coulomb interaction.
On the other hand, further increase of the Coulomb interaction leads to
quite different behavior: as the number of orbitals $M$ increases,
 the mass is less renormalized, which makes  the metallic state more
stable up to large $U$.\cite{Lu94,Fresard:SlaveBoson,Florens02} 
This implies that  the inter-orbital interaction
enhances orbital fluctuations for large $U$, which in turn stabilize 
the metallic state, as pointed out in the two-orbital case.\cite{Koga02}
In fact, the critical points $U_{c2}(T=0)=9.2173, 12.6044$ and $15.9958$ 
for $M=2, 3$ and 4 (see Appendix),
are much larger than $U_c=5.84$ for the single-orbital model.\cite{SFA}
Therefore, enhanced orbital fluctuations play a key role to
 stabilize the metallic state at zero temperature.

\begin{figure}[htb]
\includegraphics[width=\linewidth]{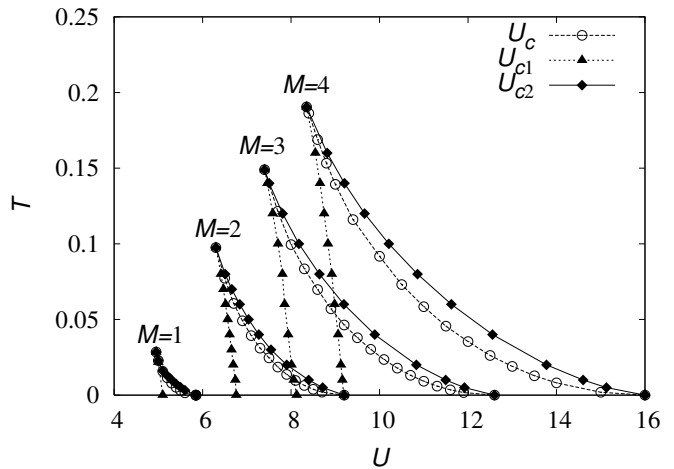}
\caption{Phase diagrams for the degenerate Hubbard model with 
the orbital degeneracy $M=1, 2, 3, 4$. }
\label{fig:phase}
\end{figure}

We now move to the finite-temperature properties. The phase
diagrams obtained in the way mentioned
in the previous section are shown in Fig. \ref{fig:phase}.
%%%%%%%%%%%%%%%%%%%%%%%%%%%%%%
%%%%%%%%%%%%%%%%%%%%%%
There are three phase boundaries for each system,
 $U_{c1}$, $U_{c2}$  and $ U_c$.
The double-well structure in the grand potential
gives rise to two kinds of transitions with the discontinuity 
in the physical quantities.
On the phase boundary $U_{c1} (U_{c2})$, 
the Mott transition occurs from the insulating (metallic) state 
to the metallic (insulating) state as $U$ decreases (increases).
The area surrounded by $U_{c1}$ and $ U_{c2}$ is 
referred to as the coexistence phase.\cite{GeorgesRev,SFA}
The critical point $U_c$ is determined so that the two minima 
of the grand potential take the 
same value. At zero temperature,
the critical points $U_{c}$ and $U_{c2}$ merge to give the continuous 
Mott transition, because the double-well structure disappears at $T=0$. 
Therefore, the introduction of the temperature drastically changes 
the phase boundary as discussed in the single-orbital case. \cite{GeorgesRev,SFA}
%%since the grand potential for the Mott insulating state
%%is decreased linearly due to the residual entropy at zero temperature.
It is found that as temperature increases, the phase boundaries 
$U_{c1}$, $U_{c2}$, and $U_c$ merge at the critical temperature $T_c$,
where the second-order phase transition occurs.

\begin{figure}[htb]
\includegraphics[width=\linewidth]{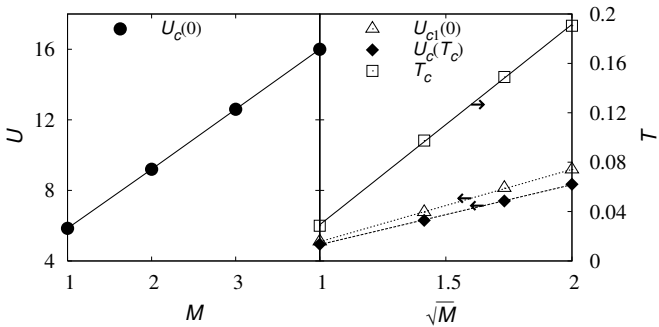}
\caption{The critical points $U_c(T=0)$, $U_c(T=T_c)$, $U_{c1}(T\rightarrow 0)$ and  the critical temperature $T_c$ as a function of the orbital 
degeneracy $M$.
}\label{MvsUc}
\end{figure}

The above characteristic properties in the Mott
transitions are qualitatively the same as the single-orbital case.
We now discuss the effects due to orbital fluctuations quantitatively.
First note that the coexistence region bounded by the phase boundaries 
$U_{c1}$ and $U_{c2}$ is enlarged when the number of orbital $M$ 
 increases.  Furthermore, we see that
the $M$-dependence of the critical points $U_{c1}$, $U_{c}(T_c)$ and 
the critical temperature $T_c$ is different from that of $U_{c2}$.
To clarify this point, 
the critical values are plotted as a function of $M$ in Fig. \ref{MvsUc}.
%%%%%%%%%%%%%%%%%%%%%%%%%
%%%%%%%%%%%%%%%%%%%%%%%%
It is found that $U_{c2}$ is proportional to the orbital degeneracy $M$,
while $U_{c1}$ and $U_c(T_c)$ the square root of $M$.
These results are consistent with the previous results.
\cite{Lu94,Fresard:SlaveBoson,Hasegawa98,Florens02,Ono03}
However, we wish to note that the finite-temperature properties 
such as the phase diagrams have not been clarified 
quantitatively so far.
In particular, the results on the $M$-dependence of 
the critical temperature $T_c$ obtained here are beyond
the qualitative discussions by Florens {\it et al.},\cite{Florens02} 
who obtained  only the upper bound for $T_c \propto M$.

%At higher temperature the transition point shows the same dependency as $U_c(T_c)$. Therefore we conclude that in finite temperature $U_c(T)$ shows square root dependency on $M$ rather linear dependency which was suggested in Refs. \cite{Gunnarsson96,Han98,Koch99}. 

%%%%%%%%%%%%%%%%%%%%%%%%%
\begin{figure}[htb]
\includegraphics[width=\linewidth]{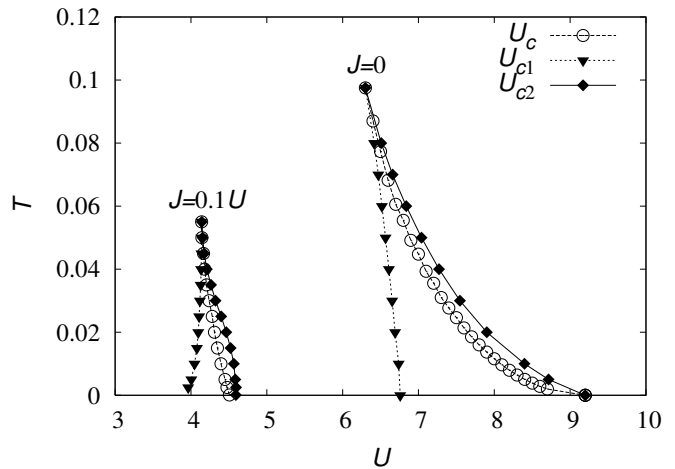}
\caption{
Phase diagrams for the two-orbital Hubbard model with 
the Hund coupling $J=0.1U$ under the condition $U=U'+2J$.
For clarity, we also show the results at $J=0$.
}\label{hund}
\end{figure}
%%%%%%%%%%%%%%%%%%%%%%

To conclude this section, we would like to briefly discuss the effects 
of the Hund exchange coupling.  For this purpose, we add the 
following term to the Hamiltonian (\ref{eq:original_model}),
%%%%%%%%%%%%%%%%%%%%%%%%%%%%%
\begin{eqnarray}
{\cal H}_J &=&-   J \sum_i\sum_{\alpha<\alpha^\prime}
              \sum_{\sigma }
                n_{i \alpha \sigma} n_{i \alpha^\prime \sigma} \nonumber\\
          &&  -   J \sum_i\sum_{\alpha<\alpha^\prime}
                (c^\dag_{i \alpha \uparrow}c_{i \alpha \downarrow}
                c^\dag_{i \alpha^\prime \downarrow}c_{i \alpha^\prime \uparrow}+H.c.)\nonumber\\
          &&  -   J \sum_i\sum_{\alpha<\alpha^\prime}
                (c^\dag_{i \alpha \uparrow}c^\dag_{i \alpha \downarrow}
                c_{i \alpha^\prime \uparrow}c_{i \alpha^\prime \downarrow}+H.c.)
%         &-&   J \sum_{\alpha<\alpha^\prime=1}^M
%               (c^\dag_{i \alpha \uparrow}c_{i \alpha \downarrow}
%                c^\dag_{i \alpha^\prime \downarrow}c_{i \alpha^\prime \uparrow}+H.c.)
                \label{eq:hund}
\end{eqnarray}
%%%%%%%%%%%%%%%%%%%%%%%%%%%%%
with $J>0$,
where we impose the condition $U=U'+2J$ due to the symmetry 
requirement.  Since the calculation gets somewhat difficult
in the presence of $J$, we show the results obtained 
for the representative case of $M=2$.  The obtained phase 
diagram is shown in Fig. \ref{hund} and compared with the $J$ =0 case.
First, we notice that upon introducing $J$, the phase boundaries
are immediately shifted to the weak-interaction regime, and 
therefore the metallic state gets unstable for large $U$. 
Correspondingly, the coexistence region surrounded by the first order transitions
 shrinks as $J$ increases.  This remarkable tendency indeed
reflects the fact that the metallic state is stabilized by 
enhanced orbital fluctuations  in the case of $U=U'$: the Hund 
coupling suppresses such orbital fluctuations, and 
stabilizes the Mott-insulating phase.
 Another point to be mentioned is that the Mott transition becomes of
first-order even at zero temperature in the presence of
$J$,\cite{Bunemann:Gutzwiller,Ono03,Pruschke04} since the subtle balance realized at $T=0$ in the 
case of $U=U'$ is not kept anymore for finite $J$.
Since there is another claim that the second order transition is possible for certain 
choices of parameters at $T=0$,\cite{Pruschke04} which may depend on
the strength of Hund coupling, we need further detailed discussions to clarify this problem.
Although we have presented only the results for the 
 $M=2$ case, we expect that the effects of $J$ on the phase diagram
should be essentially the same as shown 
here: the introduction of $J$ suppresses
orbital fluctuations, thus favors the Mott insulating
phase even in the small $U$ regime. It also shrinks the 
coexistence phase dramatically.

%%%%%%%%%%%%%%%%%%%%%%%%%%%%%%%%%%%%%%%%%%%%%%%%%%%%%%%%%%%%%%%%%%%%%%%%%%%
%     DISCUSSION for SQUARE ROOT of M
%%%%%%%%%%%%%%%%%%%%%%%%%%%%%%%%%%%%%%%%%%%%%%%%%%%%%%%%%%%%%%%%%%%%%%%%%%%
%This may be explained by the simple estimation of the Hubbard gap.
%In the Mott insulating phase, the Hubbard gap is obtained from the 
%second moment of the hopping Hamiltonian for a state $|\psi\rangle$
%with a extra electron (hole). All possible configurations for electrons
%have the same energy. Therefore, the effective hopping matrix 
%$T^2=\langle \psi | H^2 |\psi \rangle$ is given as
%$T=\sqrt{M}t$. Therefore the Hubbard gap $\Delta=U-2zT$ depends on 
%the square root of the orbital degeneracy $M$.
%%%%%%%%%%%%%%%%%%%%%%%%%%%%%%%%%%%%%%%%%%%%%%%%%%%%%%%%%%%%%%%%%%%%%%%%%%%

%%%%%%%%%%%%%%%%%%%%%%%%%%%%%%%%%%%%%%%%%%%%%%%%%%%%%%%%%%%%%%%%%%%%%
%%%                           conclusion                           %%
%%%%%%%%%%%%%%%%%%%%%%%%%%%%%%%%%%%%%%%%%%%%%%%%%%%%%%%%%%%%%%%%%%%%%

\section{Summary}\label{sec:summary}

We have investigated the Mott transitions in the multi-orbital
Hubbard model at zero and finite temperatures 
by means of the self-energy functional approach. 
By choosing the Anderson impurity model as a reference system,
we have discussed how the orbital degeneracy affects 
the nature of the Mott transitions.
We have obtained the finite-temperature phase diagram for
the system having $M$($\le 4$) degenerate bands.
Although the phase diagrams show  qualitatively similar
features irrespective of the orbital degeneracy, there
are some remarkable effects due to degenerate orbitals.
In particular, it has been shown that enhanced orbital fluctuations
make the metallic phase more stable even at finite temperatures.
Therefore, if such fluctuations are suppressed, the metallic
state is expected to be unstable, which has been  shown to be
the case by considering the system with the Hund coupling.
Also, it has quantitatively been clarified how
distinctly  the $M$-dependence  appears 
in the  critical points $U_{c1}$ and $U_{c2}$:
the critical point $U_{c1}$
depends on the square root of  the orbital degeneracy $M$, while
the critical point $U_{c2}$ is proportional to $M$.
%The end point of first order transition $T_c$ depends on $\sqrt{M}$ and 
%$U_c(\sim T_c)$ shows the same dependence. 
%So $U_c$ at finite temperature shows square root dependence rather linear 
%dependence. 
This analysis concludes that the critical temperature $T_c$ 
is proportional to the square root of $M$, 
which may be important to understand the Mott transitions 
in the real materials with orbital degeneracy.

The self-energy functional approach used in the paper 
allows quantitatively reliable discussions 
for multi-orbital systems at finite temperatures.
Since this formalism provides a tractable way to incorporate 
spin and charge ordered states, which have been neglected in 
this paper, it may be used for
more detailed study of the finite-temperature properties in
multi-orbital Mott systems.

%%%%%%%%%%%%%%%%%%%%%%%%%%%%%%%%%%%%%%%%%%%%%%%%%%%%%%%%%%%%%%%%%%%%%
%%%                         acknowledgments                        %%
%%%%%%%%%%%%%%%%%%%%%%%%%%%%%%%%%%%%%%%%%%%%%%%%%%%%%%%%%%%%%%%%%%%%%
\begin{acknowledgments}
Numerical computations were carried out at the Supercomputer Center, 
the Institute for Solid State Physics, University of Tokyo. 
This work was supported by a Grant-in-Aid for Scientific Research from 
the Ministry of Education, Culture, Sports, Science, and Technology, Japan.
\end{acknowledgments}

%%%%%%%%%%%%%%%%%%%%%%%%%%%%%%%%%%%%%%%%%%%%%%%%%%%%%%%%%%%%%%%%%%%%%
%%%                            Appendix                            %%
%%%%%%%%%%%%%%%%%%%%%%%%%%%%%%%%%%%%%%%%%%%%%%%%%%%%%%%%%%%%%%%%%%%%%
\appendix
\section{}
We analytically determine the critical value of $U$ for the
 Mott transition in the multi-orbital Hubbard model 
at zero temperature.
The grand potential Eq. (\ref{eq:Omega_DIA}) is rewritten as
%%%%%%%%%%%%%%%%%%%%%%%%%%%%%%%
\begin{eqnarray}
\Omega /L=\Omega_{\rm imp}
%    &+&2M T\sum _i \int dz \rho(z)  F(\omega_i(z)) \nonumber\\
%    &-&2M T\sum_i F( \omega_i), \nonumber\\
%      =\Omega^\prime
    &+&2M \sum _i \int dz \rho(z)  
    \omega_i(z)\theta \left[-\omega_i(z)\right] \nonumber\\
    &-&2M \sum_i \omega_i \theta(-\omega^\prime_i),\label{eq:Omega_zeroT} 
\end{eqnarray}
%%%%%%%%%%%%%%%%%%%%
where $\Omega_{\rm imp}$ is the grand potential of the
reference system and 
%%$F(x)=-T \ln (1+\exp(x/T))$, 
$\theta(x)$ is the step function.  $\omega^\prime_i, \omega_i(z)$ 
are the poles of the impurity Green's function of the reference 
system, $G^\prime(\omega)$, and the approximated Green function 
of the original system, $G(\omega;z)=1/(\omega +\mu - z - \Sigma(\omega))$, 
respectively.

First we consider the two-orbital system. 
In the atomic limit $V=0$,
the Green's function and the self-energy at the impurity site for the  
reference system are given by
%%%%%%%%%%%%%%%%%%%%%
\begin{eqnarray}
G^\prime_{\rm atom}(\omega)&=&\frac{1}{2}\left[\frac{1}{\omega-U/2}+\frac{1}{\omega+U/2}\right]\\
\Sigma_{\rm atom} (\omega)&=&\mu+
%\frac{1}{\omega}\left(\frac{U}{2}\right)^2.
\frac{U^2}{4\omega},
\end{eqnarray}
%%%%%%%%%%%%%%%%%%%%%
where the poles of $G^\prime_{\rm atom}$ are $\pm U/2$.
Then the Green's function of the original lattice model is given by
%%%%%%%%%%%%%%%%%%%%
\begin{eqnarray}
G^{-1}_{\rm atom}
%^{\rm atom}
(\omega;z)
%        &=&\omega +\mu - z - \Sigma_\alpha(\omega)\nonumber\\
        &=&\omega - z + \frac{U^2}{4\omega},
\end{eqnarray}
%%%%%%%%%%%%%%%%%%%%
which has the poles  at
%%%%%%%%%%%%%%%%%%%%%%%%%%%%%%%%%%%%%
\begin{equation}
\omega_{\pm}(z)=\frac{1}{2}\left(z \pm \sqrt{z^2+U^2}\right).
\label{eq:poles_Gwz}
\end{equation}
%%%%%%%%%%%%%%%%%%%%%%%%%%%%%%%

As discussed in the text, 
we can expand the quantities in $V$ around the atomic limit.
The grand potential is $\Omega^\prime =-2U-24V^2/U+{\cal O}(V^4)$.
We also analyze the Green's functions $G^\prime(\omega)$ and $G(\omega;z)$ 
around the atomic limit and obtain their poles up to 
the second order in $V$.
Here we need only the poles in the negative energy region, because 
the poles 
in the  positive energy region do not affect $\Omega$, see 
Eq. (\ref{eq:Omega_zeroT}).  The
Green's functions $G^\prime(\omega)$ and $G(\omega;z)$ have eight 
poles in the $M=2$ case, and thus four poles in the negative region.
%and a free green function for the lattice model is 
%$g_0^{-1}(\omega)=\omega + \mu -V^2/\omega$, 
%The Green function for the lattice model is written by
%\begin{equation}
%G(\omega)=\sum_{i=1}^{n_{p}} \frac{R_i}{\omega-\varepsilon_i}, \label{eq:Green_funtion}
%\end{equation}
%where $n_p(=8)$ is the number of poles of $G(\omega)$ and 
The negative poles of $G^\prime(\omega)$ are
%%%%%%%%%%%%%%%%%%%%%
\begin{eqnarray}
\omega^\prime _{1}&=&- 10 \frac{V^2}{U}+{\cal O}(V^4),\nonumber\\
\omega^\prime _{2}&=&-\frac{U}{2}-(30+4\sqrt{3})\frac{V^2}{U}+{\cal O}(V^4),
\nonumber\\
\omega^\prime _{3}&=&-\frac{U}{2}-(30-4\sqrt{3})\frac{V^2}{U}+{\cal O}(V^4),
\nonumber\\
\omega^\prime _{4}&=&-2U-26\frac{V^2}{U}+{\cal O}(V^4),\nonumber
\end{eqnarray}
%and
%\begin{eqnarray}
%R_{1(5)}&=&50 \frac{V^2}{U^2}+{\cal O}(V^4),\nonumber\\
%R_{2(6)}&=&\frac{1}{4}-\frac{1}{3}(79-19 \sqrt{3})\frac{V^2}{U^2}+{\cal O}(V^4),\nonumber\\
%R_{3(7)}&=&\frac{1}{4}-\frac{1}{3}(79+19 \sqrt{3})\frac{V^2}{U^2}+{\cal O}(V^4),\nonumber\\
%R_{4(8)}&=&\frac{8}{3}\frac{V^2}{U^2}+{\cal O}(V^4).\nonumber
%\end{eqnarray}
%Now we need only negative poles $\{\varepsilon_i\} (i=1,2,3,4)$, see Eq. \ref{eq:Omega_zeroT}. Considering the limit $V \to 0$, because a residue disappear  $R_{1(4)}\to 0$ accompanying with that poles merging with zeros and both of them disappear, the zeros $\{\omega_i\}$ of $G(\omega)$ is given by $\omega_{2(4)}\to 0^-,(-2U)$. $\varepsilon_2$ and $\varepsilon_3$ merge at the limit of $V\to 0$, this leads $\omega_3\to -U/2$. There is a fictitious zero  $\omega_1=0$, due to zero of $g=1/(\omega+\mu-V^2/\omega)$. $\omega_1$ is canceled out after subtracting $g^{-1}-G^{-1}$. So there are three negative poles of $\Sigma(\omega)$ in the limit of $V\to 0$, $\{\varepsilon^\Sigma_i\}=0^-,-U/2,-2U$. 

%Finally we consider poles of $G(\omega;z)$. In the atomic limit there is only one negative pole $\varepsilon_-(z)$ eq. (\ref{eq:poles_Gwz}). $G(\omega;z)$ should have same number of pole as $G(\omega)$. Three negative poles merge with negative poles of $\Sigma(\omega)$ and disappear at $V=0$. Thereby we can get four 
%%%%%%%%%%%%%%%%%%%%%%%%%%%%%
and the negative poles of $G(\omega;z)$ are
\begin{eqnarray}
\omega _1(z)&=&100zV^2/U^2 +{\cal O}(V^4),\nonumber\\
\omega _2(z)&=&\omega _{-}(z)+B(z/U)\ V^2/U+{\cal O}(V^4) ,\nonumber\\
\omega _3(z)&=&-U/2-30V^2/U+{\cal O}(V^4) ,\nonumber\\
\omega _4(z)&=&-2U-\frac{168z+390U}{15U+8z}V^2/U+{\cal O}(V^4), \nonumber\\
\end{eqnarray}
where
\begin{eqnarray}
B(x)=-\frac{6525+8714x^2-3200x^4}{(225-64x^2)\sqrt{1+x^2}}
-\frac{50x(237-64x^2)}{225-64x^2}.\nonumber\\
\end{eqnarray}
The grand potential is written down as
\begin{eqnarray}
\Omega/L&=&-4\int^{\infty}_{-\infty}dz\rho(z)\xi(z,U)\nonumber\\
&+&\frac{4V^2}{U^2}\bigg\{39U+100\int_{-\infty}^0dz \rho(z)
\left(z-\xi(z,U)\right)\nonumber\\
 &+&U\int^\infty_{-\infty}dz \rho(z)\frac{317U+83 
\xi(z,U)}{[17U+8\xi(z,U)]\xi(z,U)}
\bigg\}+{\cal O}(V^4),\nonumber\\
 \label{eq:Omega_perturbation}
\end{eqnarray}
where $\xi(z,U)=\sqrt{U^2+z^2}$.
The condition $\partial^2 \Omega/\partial V^2|_{V=0}= 0$ is 
satisfied at the Mott transition point $U_c$.\cite{SFA} Given 
that the free density of state is symmetric $\rho(z)=\rho(-z)$, 
we derive the self-consistent equation for $U_c$, Eq. (\ref{eq:SCEq_Uc}). 

Similarly, we can analyze the cases of $M=3$ and $M=4$.
The self-consistent equation for $U_c$ at $M=3$ reads
%%%%%%%%%%%%%%%%%%%%%%%%%%%%%%%%%%%
\begin{eqnarray}
U_c&=&-\frac{49}{20}\int^0_{-\infty} \left[z-\xi(z,U_c)\right)] 
\rho (z)dz\nonumber\\
&&-\frac{U_c}{80}\int^\infty_{-\infty}\frac{545U_c+80 
\xi(z,U_c)}{[17U_c+8 \xi(z,U_c)]\xi(z,U_c)}
\rho(z)dz,\nonumber\\
\label{eq:M3:SCEq_Uc}
\end{eqnarray}
%%%%%%%%%%%%%%%%%%%%%%%%%%%
and at $M=4$ is
\begin{eqnarray}
U_c&=&-\frac{324}{135}\int^0_{-\infty} \left[z-\xi(z,U_c)\right] 
\rho (z)dz\nonumber\\
&&-\frac{U_c}{135}\int^\infty_{-\infty}\frac{837U_c+63  
\xi(z,U_c)}{[17U_c+8 \xi(z,U_c)]
\xi(z,U_c)}\rho(z)dz.\nonumber\\
\label{eq:M4:SCEq_Uc}
\end{eqnarray}
%%%%%%%%%%%%%%%%%%%%
The solution of these equations gives the accurate values of the
critical point:
$U_{c}=12.6044$ and $15.9958$ for $M=3$ and $4$, respectively.

\newpage %Just because of unusual number of tables stacked at end

%%\bibliography{main}
%%%%%%%%%%%%%%%%%%%%%%%%%%%%
% REFERENCES
%%%%%%%%%%%%%%%%%%%%%%%%%%%%
%\bibliography{main}

\end{document}